\documentclass[fleqn,twoside]{article}
\usepackage[headings]{espcrc2}
\readRCS
$Id: espcrc2.tex,v 1.2 2004/02/24 11:22:11 spepping Exp $
\usepackage{graphicx}

\newcommand{\AmS}{{\protect\the\textfont2
  A\kern-.1667em\lower.5ex\hbox{M}\kern-.125emS}}
\def\eps{\epsilon}
\def\as{\alpha_s} 
\def\nn{\nonumber}

\title{The sector decomposition approach to real radiation at NNLO}

\author{G. Heinrich\address{Institut f\"ur Theoretische Physik, 
Universit\"at Z\"urich,
Winterthurerstrasse 190, 8057 Z\"urich, Switzerland}
\thanks{
 To appear in the proceedings of the 
 7th International Symposium on Radiative Corrections (RADCOR05), 
 Shonan Village, Japan, 2005.}
 } 
\runtitle{}
\runauthor{}

\begin{document}

\begin{abstract}
A method based on sector decomposition is briefly described
which is suitable for the calculation of 
the double real radiation part of $e^+e^- \to$ 3 jets at 
next-to-next-to-leading order  in a fully differential way.
\vspace{1pc}
\end{abstract}

\maketitle
\thispagestyle{myheadings}

\markright{ZU-TH 06/06}

\section{Introduction}

The successful comparison with theory of a wealth of data collected at high energy collider
experiments in the past decades largely relied on the calculation of radiative 
corrections. Calculating at next-to-leading order (NLO) accuracy 
in perturbation theory is in many cases sufficient to match the experimental 
precision. However, there are also prominent exceptions  
where the NLO 
predictions still suffer from large  uncertainties,  
or where the experimental precision is extremely high. 
The latter is the case for measurements of jet rates and shape observables 
in $e^+e^-$ annihilation, which for example allow for a precise determination of 
the strong coupling constant $\alpha_s$. 
As only NLO predictions are available for $e^+e^-\to 3$\,jets, 
the corresponding LEP measurements 
have not been included in the current world average for 
$\alpha_s$.
A future International Linear Collider (ILC) will allow for precision
measurements at the per-mille level, which offer the possibility 
of a determination of  $\alpha_s$ 
with unprecedented precision. 
However, this will only be possible if theoretical predictions 
for jet rates and related observables at NNLO accuracy are available.

The calculation of $e^+e^-\to  3$\,jets at order $\as^3$ 
requires the calculation of virtual two-loop corrections 
combined with a $1\to 3$ parton phase space, 
one-loop corrections combined with a $1\to 4$ parton phase space
where one parton can become soft and/or collinear (``theoretically unresolved"), 
and the tree level matrix element
squared for $1\to 5$ partons where up to two partons 
can become  unresolved. 
After the virtual two-loop corrections 
have become available\,\cite{Garland:2001tf}, 
the stumbling block now is given by the double real radiation part, 
where the unresolved particles lead to a complicated infrared 
singularity structure which manifests itself 
upon phase space integration. 
These singularities have to be subtracted and cancelled with the ones 
from the virtual contributions before a Monte Carlo program 
can be constructed. 
To achieve this task, 
two different approaches have been followed, 
one relying on the manual construction of an analytical subtraction 
scheme\,\cite{Kosower:2002su,Weinzierl:2003fx,Gehrmann-DeRidder:2004tv,Kilgore:2004ty,Frixione:2004is,Gehrmann-DeRidder:2004xe,Somogyi:2005xz,Gehrmann-DeRidder:2005cm}, 
the other one relying on sector 
decomposition\,\cite{Heinrich:2002rc,Gehrmann-DeRidder:2003bm,Anastasiou:2003gr,Binoth:2004jv,Heinrich:2004jv,Anastasiou:2004qd,Anastasiou:2005qj,Anastasiou:2005pn,Heinrich:2006sw}.
The application of sector decomposition\,\cite{Hepp:1966eg,Roth:1996pd,Binoth:2000ps} 
to NNLO phase space integrals 
has first been proposed in \cite{Heinrich:2002rc}. Subsequently, it has been 
applied to calculate inclusive phase space integrals for 
$e^+e^-\to 2$\,jets at NNLO\,\cite{Gehrmann-DeRidder:2003bm,Binoth:2004jv,Heinrich:2004jv}.
The combination of the sector decomposition approach with a 
measurement function  first has been presented in \cite{Anastasiou:2003gr},  
and already lead to a number of important results\,
\cite{Anastasiou:2004qd,Anastasiou:2005qj,Anastasiou:2005pn}.
Its application to the double real radiation part of 
$e^+e^-\to  3$\,jets at order $\as^3$\,\cite{Heinrich:2006sw} 
represents a new degree of complexity and therefore will be a 
crucial test of the potential of this method. 


The main features of the two approaches are the following: 
Within the methods based on the explicit construction  of a 
subtraction scheme, 
the subtraction terms are integrated analytically  over the 
unresolved phase space 
such that the pole coefficients are obtained in analytic form. 
This requires appropriate phase space factorisation and subtraction 
terms which are simple enough to be integrated 
analytically in $D=4-2\eps$ dimensions.
The method naturally leads to 
a close to minimal number of subtraction terms, and
observing the way how the poles of the different contributions 
cancel allows insights into the infrared structure of QCD.

In the sector decomposition approach, the poles are isolated 
by an automated routine and the pole coefficients are integrated 
numerically. The advantages of this approach reside in the fact that 
the extraction of the infrared poles is algorithmic, 
and that the subtraction terms can be arbitrarily 
complicated as they are integrated only numerically.  On the other
hand, the algorithm which isolates the poles 
increases the number of original functions
and in general does not lead to the 
minimal number of subtraction terms, 
thus producing rather large expressions.

\section{The Method}
The universal applicability of sector decomposition goes back 
to the fact that it acts in parameter space by a simple mechanism. 
The parameters can be Feynman parameters in the case of multi-loop 
integrals, or phase space integration variables, or a combination of both. 
In the following, the working mechanism of sector decomposition
will be outlined only briefly, details can be found 
in \cite{Binoth:2000ps,Heinrich:2006sw}.

An overlapping singularity in parameter space is of the type
\begin{eqnarray}
I&=&
\int_0^1 dx\int_0^1 dy \,x^{-1-\epsilon}\,(x+y)^{-1}\;,
\label{ol}
\end{eqnarray}
where a naive subtraction of the singularity for $x\to 0$ 
fails. 
To solve this problem, one can split the integration region into sectors 
where the variables $x$ and $y$ are ordered by multiplying (\ref{ol}) 
with unity in the form 
$[\underbrace{\Theta(x-y)}_{(a)}+\underbrace{\Theta(y-x)}_{(b)}]$. 
Then the integration domain  is remapped to the unit cube: 
after the substitutions $y=x\,t$ in sector (a) and 
$x=y\,t$ in sector (b), one has
\begin{eqnarray}
I&=&\int_0^1 dx\,x^{-1-\epsilon}\int_0^1 dt
\,(1+t)^{-1}\nn\\
&&+\int_0^1 dy
\,y^{-1-\epsilon}\int_0^1 dt\,t^{-1-\epsilon}\,(1+t)^{-1}\;,
\end{eqnarray}
where the singularities are now factorised. 
For more complicated functions, several iterations 
of this procedure may be necessary, but it is easily implemented 
into an automated subroutine. Once all singularities are factored out, 
they can be subtracted  and 
the result can subsequently be expanded in $\epsilon$. Note that  
the subtractions of the pole terms 
naturally lead to plus distributions\,\cite{Anastasiou:2003gr} by the identity 
\begin{eqnarray*}
x^{-1+\kappa\epsilon}=\frac{1}{\kappa\,
\epsilon}\,\delta(x)+
\sum_{n=0}^{\infty}\frac{(\kappa\epsilon)^n}{n!}
\,\left[\frac{\ln^n(x)}{x}\right]_+\,,\,\mbox{where}
\end{eqnarray*}
\begin{eqnarray*}
 \int_0^1 dx \, f(x)\,\left[g(x)/x\right]_+=\int_0^1 dx \, 
\frac{f(x)-f(0)}{x}\,g(x)\;.
\end{eqnarray*}
In this way, a 
Laurent series in $\eps$ is obtained, where the pole coefficients are 
sums of finite parameter integrals which can be evaluated numerically. 

For the numerical evaluation of loop integrals it has to be assured that 
no integrable singularities 
are crossed which spoil the numerical convergence. 
For integrals depending only on a single scale, which can be factored out, 
this does not pose a problem at all. 
For integrals with more than one scale, like for example two-loop box 
diagrams, the situation is more difficult, but 
in the case of $e^+e^-$ annihilation to massless final state 
particles, 
evaluation over the whole physical region is possible,  
as the kinematics of these 
processes is such that the Mandelstam variables are always non-negative. 

\section{Application to $e^+e^-\to 3$\,jets at NNLO}
In order to focus on a concrete example of phenomenological relevance, 
we will discuss the application of sector decomposition to the 
calculation of $e^+e^-\to 3$\,jets at NNLO in the following. 
Although the virtual contributions are not the main subject of this talk, 
they will be commented on briefly.

\subsection{Double virtual and mixed real-virtual contributions}

The contributions to the amplitude which involve virtual integrals are composed of 
the two-loop (and one-loop times one-loop) 
corrections combined with a $1\to 3$ particle phase space, 
and the mixed real-virtual contributions 
where one-loop corrections are combined with a $1\to 4$ particle phase space 
with up to one unresolved particle. 
In both cases, sector decomposition for loop integrals\,\cite{Binoth:2000ps} 
can serve to extract the poles in $1/\eps$. 

In the case of the two-loop integrals, the result will only 
depend on the invariants $y_1=s_{12}/q^2, y_2=s_{13}/q^2$ and $y_3=s_{23}/q^2$, 
where $q^2$ is the invariant mass of the $e^+e^-$ system and 
$\sum_{i=1}^3 y_i=1$. 
However, as the full two-loop matrix element is known 
analytically\,\cite{Garland:2001tf}, 
this part could also be taken from the literature, 
thus gaining a considerable amount of CPU time.
The subsequent phase space integration over the $y_i$
is tri\-vial, and the 3-jet measurement function will make sure that 
all events where a singular limit $y_i\to 0$ is approached will be rejected. 

In the case of the one-loop contributions, the most complicated 
objects will be 5-point integrals with one off-shell external leg. 
Sector decomposition will lead to a result in terms of five 
independent scaled Mandelstam invariants $y_i$. 
This result has to be 
calculated up to order $\eps^2$, as it will be combined with 
the  $1\to 4$ parton phase space where one parton can become unresolved, 
leading to $1/\eps^2$ poles. This does in principle not constitute a 
problem, as the expansions to higher order in $\epsilon$, 
as well as  $1\to 4$ parton phase space integrals,  
are well under control within sector 
decomposition. 
It is also possible to do parts of the loop integrations analytically to achieve 
a form which is suitable for subsequent 
sector decomposition\,\cite{Heinrich:2004jv,Anastasiou:2005pn}.
However, these contributions have not yet been implemented 
into the Monte Carlo program.

\subsection{Real radiation at NNLO}

\begin{figure}[htb]
\includegraphics[height=4.5cm]{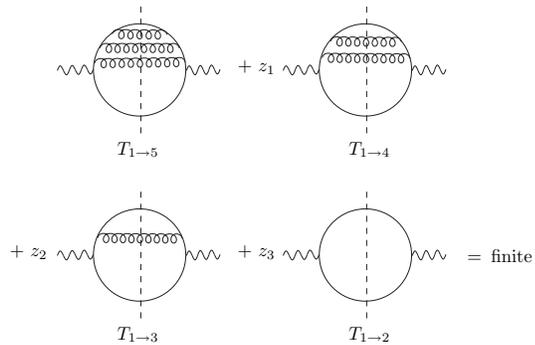}
\vspace*{-7mm}
\caption{Cancellation of IR divergences in the sum over all cuts 
of the renormalised graph}
\label{fig2}
\end{figure}

In \cite{Heinrich:2006sw}, the method based on sector decomposition is 
for the first time applied to extract the poles appearing in massless $1\to 5$ 
particle integrals. 
The correctness of the results for the integrals 
over the $1\to 5$ particle phase space can be checked by  exploiting  
the fact that the sum over all cuts of a given 
(UV renormalised) diagram must be infrared finite. 
This is shown in Figure~\ref{fig2} for a sample diagram:  
Summing over all cuts of this diagram and performing UV renormalisation, 
we obtain the condition 
\begin{eqnarray}
T_{1\to5}+z_1\,T_{1\to 4}+z_2\,T_{1\to 3}+z_3\,T_{1\to 2}=\mbox{finite}\;,
\label{cfin}
\end{eqnarray}
where $T_{1\to i}$ denotes the diagram with $i$ cut lines.

\vspace*{-6mm}

\vspace*{-5mm}

The renormalisation constants $z_i$ (in Feynman gauge) are 
given by\,\cite{Binoth:2004jv,Heinrich:2006sw} 
\begin{eqnarray}
z_1&=&C_F\frac{\as}{4\pi}\,\frac{1}{\eps}\, ,\, 
z_2=C_F^2\left(\frac{\as}{4\pi}\right)^2\,
\left(\frac{1}{2\eps^2}-\frac{1}{4\eps}\right)\nonumber\\
z_3&=&C_F^3\left(\frac{\as}{4\pi}\right)^3\,
\left(\frac{1}{6\eps^3}-\frac{1}{4\eps^2}+\frac{1}{6\eps}\right)\;.
\end{eqnarray}
The important new ingredient 
in  eq.~(\ref{cfin}) is the calculation of $T_{1\to 5}$. 
The sector decomposition method leads to \cite{Heinrich:2006sw}
\begin{eqnarray}
T_{1\to 5}
&=&-C_F^3\left(\frac{\as}{4\pi}\right)^3\,T_{1\to 2}
\left\{
\frac{0.16662}{\eps^3}\right.\nonumber\\
&&\left.+\frac{1}{\eps^2}\,
[1.4993-0.4999\,\log{\left(\frac{q^2}{\mu^2}\right)}]\right.\nn\\
&&\left.+\frac{1}{\eps}\,[ 5.5959-4.4978\,
\log{\left(\frac{q^2}{\mu^2}\right)}\right.\nonumber\\
&&+\left.
0.7498\,\log^2{\left(\frac{q^2}{\mu^2}\right)} ]\,+\,\mbox{finite}
\right\}\;,
\label{t5}
\end{eqnarray}
where the numerical accuracy is  1\%. 
The expressions entering eq.~(\ref{cfin}) for $i<5$ 
combine to \cite{Heinrich:2006sw}
\begin{eqnarray}
&&z_1\,T_{1\to 4}+z_2\,T_{1\to 3}+z_3\,T_{1\to 2}=\nn\\
&&C_F^3\left(\frac{\as}{4\pi}\right)^3\,T_{1\to 2}
\left\{\frac{1}{6\eps^3}+\frac{1}{2\eps^2}
\,[3-\log{\left(\frac{q^2}{\mu^2}\right)}]\right.\nonumber\\
&&+\frac{1}{\eps}\,[5.61-\frac{9}{2}\log{\left(\frac{q^2}{\mu^2}\right)}
+\frac{3}{4}\log^2{\left(\frac{q^2}{\mu^2}\right)}]\nn\\
&&+\left.
\,\mbox{finite}\right\}\;.\label{c234}
\end{eqnarray}
We can see that  the poles in (\ref{c234}) 
are exactly cancelled by the 5-parton contribution (\ref{t5}) 
within the numerical precision.

\subsection{Differential Monte Carlo program}
Although the sector decomposition approach is considered to be 
a ``numerical method", as the pole coefficients are only calculated 
numerically, 
the isolation of the poles  is an algebraic 
procedure, leading to a set of finite functions 
for each pole coefficient as well as for the finite part. 
This feature allows the inclusion of any 
(infrared safe) measurement function, at the level of the 
final Monte Carlo program, which means that the 
subtractions and expansions in $\eps$ do {\it not} have to be 
redone each time a different observable is considered.
However, in most cases of physical relevance, 
the measurement function is such that it would prevent certain poles 
from arising at all if it were included in the $\eps$-expansion. 
For example, poles associated with a 2-jet configuration, where 3 
of the 5 final state particles become theoretically unresolved, 
would be avoided ab initio by a 3-jet measurement function. 
As extra poles are rather costly in what concerns the creation 
of a large number of terms, it would therefore be desirable not to 
produce the terms associated with the subtraction of these poles 
at all in the $\eps$-expansion.  
In short, there is a certain dilemma between maximizing the flexibility to include 
any measurement function only at the stage of the final Monte Carlo program 
and limiting the size of the produced expressions. 
For $e^+e^-\to 3$\,jets at NNLO, limiting this size is one of the most important issues. 
Therefore, this dilemma has been solved by including some ``preselection rules" 
in the $\eps$-expansion which reject configurations which will surely be 
2-jet configurations. In the example shown here, this can be achieved by introducing 
a cut parameter $y^{\rm th}$ -- which must be smaller than any possible experimental 
resolution parameter $y^{\rm cut}$ -- for the variable $s_{1345}$, as $s_{1345}\to 0$
always corresponds to a 2-jet configuration. Note that this procedure does {\it not} 
introduce any dependence on $y^{\rm th}$ in the matrix element, its only effect 
is to prevent the $\eps$-expansion subroutine from doing subtractions for  
$s_{1345}\to 0$. In this way, one can reduce the size of the expressions considerably 
without loosing any flexibility in what concerns the actual measurement function 
to be included later. In fact, the architecture of the program described in 
\cite{Heinrich:2006sw} 
is the one of a partonic event generator, such that  
fully differential information about the final state is available. 
This architecture is independent of the actual matrix element, so the toy matrix element 
used in \cite{Heinrich:2006sw} and in this article can finally be replaced by 
a different module for the full matrix element, without destroying the property
of providing the four-momenta of all final state particles.

\begin{figure}[htb]
\begin{center}
\includegraphics[height=5.7cm]{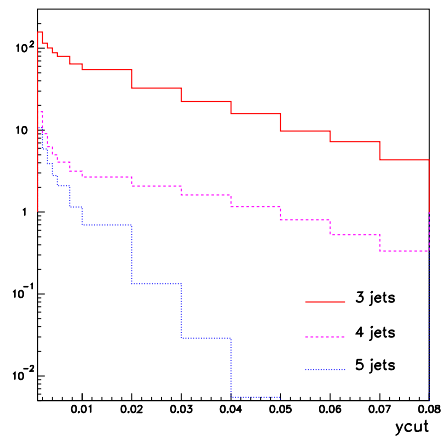}
\vspace*{-7mm}
\caption{3--, 4-- and 5--jet rates at order $\alpha_s^3$ for the toy matrix element}
\label{fig4}
\end{center}
\end{figure}

As an example, 3--, 4-- and 5--jet rates using the JADE algorithm\,\cite{Bethke:1988zc} 
are shown in Fig.~\ref{fig4}, based on a toy matrix element built from 
the graphs shown in Fig.~\ref{fig2}. 

\vspace*{-4mm}

\vspace*{-7mm}

\section{Outlook}

The sector decomposition approach to the calculation of NNLO cross sections 
is a very powerful tool, especially in what 
concerns the double real radiation part, as 
it requires neither the manual construction of subtraction terms, 
nor the factorisation of the 
phase space and the analytic integration of the subtraction  terms 
in the singular limits.  
In  \cite{Heinrich:2006sw}, the method is applied for the first time to a massless
$1\to 5$ process. The code is constructed as 
a partonic event generator, which means that fully differential information 
about the final state particles is at one's disposal.

A disadvantage of the sector decomposition approach is given by the fact that 
it produces very large expressions, as in each decomposition step, 
the number of original functions increases.
Therefore, CPU time is an issue for the treatment of  
processes with a large number of massless particles in the final state, 
as for example $e^+e^-\to 3$\,jets at  NNLO.  
However, the method sketched here  relies on a division of the amplitude squared 
into different ``topologies" corresponding to different classes of 
denominator structures, such that the problem is naturally split into smaller 
subparts. Further, the size of the expressions can be reduced by including 
information about physical limits already at the level of the $\eps$-expansion, 
without loosing any flexibility in what concerns the definition of 
observables at the Monte Carlo level.

As the method is based on a universal algorithm acting on integration 
variables,  it will surely see a number of interesting applications
in the future.  

\vspace*{3mm}

\noindent{\bf Acknowledgements}

I would like to thank the organizers of RADCOR05 for the interesting 
conference and nice atmosphere. 
This work was supported in part by the Swiss National Science Foundation
(SNF) under contract number 200020-109162.


\end{document}